\RequirePackage{fix-cm}
\documentclass[smallextended]{svjour3}       
\smartqed  
\usepackage{appendix}
\usepackage{amsmath}
\usepackage{graphicx}
\usepackage{lineno}
\usepackage{array}
\usepackage{longtable}
\usepackage[numbers]{natbib}
\usepackage{tabularx}
\usepackage{booktabs}
\usepackage{xcolor}
\usepackage{subfigure}
\usepackage[misc,geometry]{ifsym} 
\usepackage{lipsum}

\newcommand*\patchAmsMathEnvironmentForLineno[1]{%
\expandafter\let\csname old#1\expandafter\endcsname\csname #1\endcsname
\expandafter\let\csname oldend#1\expandafter\endcsname\csname end#1\endcsname
\renewenvironment{#1}%
{\linenomath\csname old#1\endcsname}%
{\csname oldend#1\endcsname\endlinenomath}}%
\newcommand*\patchBothAmsMathEnvironmentsForLineno[1]{%
\patchAmsMathEnvironmentForLineno{#1}%
\patchAmsMathEnvironmentForLineno{#1*}}%
\AtBeginDocument{%
\patchBothAmsMathEnvironmentsForLineno{equation}%
\patchBothAmsMathEnvironmentsForLineno{align}%
\patchBothAmsMathEnvironmentsForLineno{flalign}%
\patchBothAmsMathEnvironmentsForLineno{alignat}%
\patchBothAmsMathEnvironmentsForLineno{gather}%
\patchBothAmsMathEnvironmentsForLineno{multline}%
}

\begin{document}

\title{Monkeypox virus detection using pre-trained deep learning-based approaches
}


\author{Chiranjibi Sitaula \Letter  \and
        Tej Bahadur Shahi 
}

\institute{C. Sitaula  \at
              Department of Electrical and Computer Systems Engineering, Monash University\\
              Wellignton Rd, VIC 3800 \\
              \email{chiranjibi.sitaula@monash.edu}           
          \and
           TB. Shahi \at
               School of Engineering and Technology\\
               Central Queensland University \\
               Norman Garden,QLD, 4701, Australia\\
               and \\
              Central Department of Computer Science and IT \\
              Tribhuvan University\\
              TU Rd, Kirtipur 44618, Kathmandu, Nepal \\
              \email{tejshahi@cdcsit.edu.np}\\
}
\date{Received: DD Month YEAR / Accepted: DD Month YEAR}

\maketitle

\begin{abstract} 
Monkeypox virus is emerging slowly with the decline of COVID-19 virus infections around the world. People are afraid of it, thinking that it would appear as a pandemic like COVID-19. As such, it is crucial to detect them earlier before widespread community transmission. AI-based detection could help identify them at the early stage. 
In this paper, we {aim to} compare 13 different pre-trained deep learning (DL) models for the Monkeypox virus detection. For this, we initially fine-tune them with the addition of universal custom layers for all of them and {analyse the results} using four well-established measures: Precision, Recall, F1-score, and Accuracy. After the identification of the best-performing DL models, we ensemble them to improve the overall performance using a majority voting over the probabilistic outputs obtained from them. We perform our experiments on a publicly available dataset, which {results in} average Precision, Recall, F1-score, and Accuracy of 85.44\%, 85.47\%, 85.40\%, and 87.13\%, respectively {with the help of our proposed ensemble approach}. These encouraging results, which outperform the state-of-the-art methods, suggest that the proposed approach is applicable to health practitioners for mass screening.
\end{abstract}

\section{Introduction}
Monkeypox is an infectious disease caused by the
monkeypox virus (MPXV), a member of the orthopoxvirus genus. It was first identified in the monkey in 1959 at a research institute in Denmark, hence it is named as Monkeypox virus \cite{breman1980human}. Later, the first case was confirmed in humans in the Republic of Congo in 1970 when a child with smallpox-like symptoms was admitted to the hospital \cite{nolen2016extended}. It transmits to humans through close contact with infected individuals or contaminated objects \cite{reynolds2013detection}. Initially, it usually appeared in the African region but recently it has reached more than 50 countries with 3,413 confirmed cases and one death \cite{who}. Till now, there are two variants of the monkeypox virus known: one, the Central Africa clade and another, the West Africa clade.
There is no proper treatment for the monkeypox virus to date. The ultimate solution is the development of a vaccine. The diagnosis of Monkeypox can be performed mainly with the polymerase chain reaction (PCR) method or skin lesion test using electron microscopy. The most trusted method of virus confirmation is PCR, which has also been used for COVID-19 diagnosis in recent years. In addition, artificial intelligence (AI)-based techniques could help detect them with the help of virus image processing and analysis.

With the emerging growth of AI models in various domains such as chest x-ray images \cite{sitaula2021fusion}, fruit image analysis \cite{shahi2022fruit}, and sentiment analysis \cite{sitaula2021deep, shahi2022hybrid}, the AI models for medical image analysis have been proposed for various {virus-related disease detection \cite{Unnikrishnan2022, madhavan2021res}.} {For instance, Madhavan et al. \cite{madhavan2021res} developed a deep learning model (Res-COvNet) based on transfer learning approach for COVID-19 virus detection. They employed ResNet-50 \cite{he2016deep} to extract the features from X-ray images and extended the network with a classification layer. Their proposal achieved a promising accuracy of 96.2\% for identifying normal, bacterial pneumonia, viral pneumonia, and COVID-19 cases on X-ray images. Similarly, a review study on the deep learning model for COVID-19 detection was reported by Bhatt et al. \cite{Bhatt2021}. In addition, facial skin disease detection using deep learning was implemented by Yadav et al. \cite{yadav2022hsv}.}

{Besides the COVID-19 virus detection, a few works used deep learning models for other disease detection such as chicken pox, Herpes, and so on.} For instance, Sandeep et al.\cite{sandeep2022diagnosis} investigated the detection of various skin diseases such as Psoriasis, Chicken Pox, Vitiligo, Melanoma, Ringworm, Acne, Lupus, and Herpes using deep learning (DL)-based methods.  They developed a Convolutional Neural Network (CNN) to classify the skin lesion into eight disease classes and compared their solution with the help of the VGG-16 pre-trained model \cite{Simonyan15}. Their method provided an accuracy of 78\% for the detection. Low-cost image analysis for Herpes Zoster Virus (HZV) detection using CNN was proposed in \cite{lara2022low}. The early detection of HZV produced an accuracy of 89.6\% when tested on 1,000 images.
Furthermore, Measles disease detection using a transfer learning approach was implemented by Glock et al. \cite{glock2021measles}. They achieved a sensitivity of 81.7\%, specificity of 97.1\%, and accuracy of 95.2\% using the ResNet-50 model \cite{he2016deep} over the diverse rash image dataset.
Moreover, a big-data approach for Ebola virus disease detection was proposed in \cite{sarumi2020machine} using an ensemble learning approach. They utilized a combination of artificial neural network (ANN) and genetic algorithm (GA) for knowledge extraction over the big data using Apache Spark and Kafka framework.
More recently, Ahsan et. al. \cite{ahsan2022image} collected the images of Monkeypox, Chickenpox, Measles  and Normal categories using web mining techniques and verified by the experts. 
Later on, they also evaluated a transfer learning approach with the VGG-16 model considering two techniques \cite{ahsan2022monkeypox}. The first technique considered the classification of images into two disease classes: Monkeypox and Chickenpox, whereas the second technique augmented the images. 
They reported an accuracy of (97\%) while classifying the monkeypox without data augmentation, whereas the accuracy was decreased to 78\% with augmentation.

{
From existing research works on virus-related disease detection using DL methods, we observe that the majority of them have employed the transfer learning approach \cite{glock2021measles,madhavan2021res} using well-established pre-trained DL methods. Since there are not many works available on Monkeypox virus detection except the work by Ahsan et al.\cite{ahsan2022monkeypox}. Their proposal has provided encouraging results in this domain. However, it has three main limitations. First, their models only deal with binary classification with limited performance. Second, they only consider the VGG-16 DL model for transfer learning, which lacks identifying the best-performing pre-trained DL methods and their best combinations to attain optimal performance. Third, their models have insufficient interpretability. As a result, it is difficult to establish trustworthiness among health practitioners during mass screening.}

To address the aforementioned limitations, we, first, resort to the 13 pre-trained DL models and fine-tune them with the same approach. Second, we evaluate the performance of each DL model using averaged Precision, Recall, F1-score and Accuracy over 5 folds. Third, we ensemble the best-performing models to improve the overall performance. 

The main {\bf contributions} in this paper are as follows:
\begin{itemize}
    \item Propose to use a common fine-tuned architecture for all 13 pre-trained DL models for MonkeyPox detection and compare them;
    \item Perform an ablative study to select the best-performing DL models for ensemble learning;
    \item {Compare the proposed approach with the state-of-the-art methods};
    and 
    \item Show the explainability using Grad-CAM \cite{selvaraju2017grad} and LIME \cite{lime} of best-performing DL model.
\end{itemize}

\section{Materials and methods}
\label{materials_methods}

\subsection{Dataset}
Herein, we use a publicly available Monkeypox image dataset \cite{ahsan2022image,ahsan2022monkeypox}. The dataset has different sub-folders, including datasets with and without augmentations. Given that DL models prefer augmented images to learn meaningful information more accurately, we use them in this study. Table \ref{tab:dataset_description} shows the number of images from the augmented folder per category.
\begin{table}[b]
    \centering
    \caption{Dataset statistics}
    \begin{tabular}{p{1.8cm}|c|p{0.9cm}|c|p{1cm}|c}
    \toprule
   Category&Chickenpox&Measles&Monkeypox&Normal& Total\\
   \bottomrule
   \# Number&329&286&587&552&1,754\\
     \midrule
         \bottomrule
    \end{tabular}
    \label{tab:dataset_description}
\end{table}

\subsection{Evaluation metrics}
We use four widely-used performance metrics such as Precision (Eq. \eqref{eq:precision}), Recall (Eq. \eqref{eq:recall}), F1-score (Eq. \eqref{eq:f-score}), and Accuracy (Eq. \eqref{eq:acc}).

\begin{equation}
      P= \frac{TP}{TP+FP},
    \label{eq:precision}
    \end{equation}

\begin{equation}
      R= \frac{TP}{TP+FN},
    \label{eq:recall}
    \end{equation}
    
    \begin{equation}
      F= 2 \times \frac{P \times R}{P+R},
    \label{eq:f-score}
    \end{equation}
   
    \begin{equation}
    A = \frac{TP+TN}{TP+TN+FP+FN},
    \label{eq:acc}
    \end{equation}

where $TP$, $TN$, $FP$, and $FN$ represent true positive, true negative, false positive, and false negative, respectively. Similarly, $P$, $R$, $F$, and $A$ represent Precision, Recall, F1-score, and Accuracy, respectively.

\subsection{Pre-trained DL models}

The availability of various DL models trained on a large dataset, called ImageNet \cite{deng2009imagenet}, made significant progress in image classification and computer vision tasks. More precisely, when the availability of expert-labelled data is limited to some domains such as biomedical image analysis, a most common approach is to utilise these pre-trained DL models for transfer learning \cite{pan2009survey}. This is helpful to boost the performance in a limited data setting because transfer learning allows the DL models trained on large datasets to transfer learned knowledge to a small domain-specific dataset.

\begin{figure}[tbp]
	\centering
	\includegraphics[height=140mm, width=0.65\textwidth,keepaspectratio]{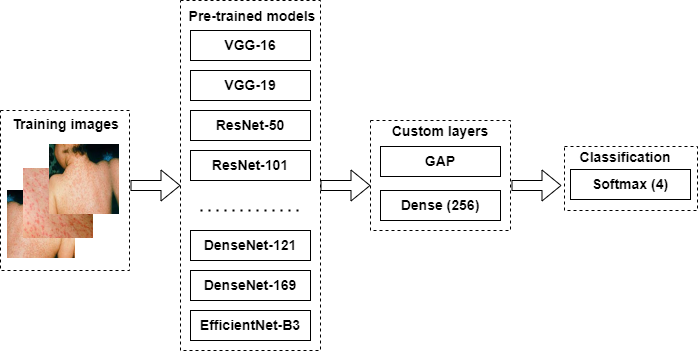}
	\caption{
 		High-level workflow to train the pre-trained DL models with custom layers. Note that GAP refers to Global Average Pooling layer and the value inside the small bracket of dense layer represents the number of units under it.
 	} 
	\label{fig:workflow}
\end{figure}

We choose 13 pre-trained DL models for this study. These pre-trained model ranges from heavy-weight DL models such as VGG-16 \cite{Simonyan15}, InceptionV3 and Xception \cite{chollet2017xception} to light-weight models such as MobileNet \cite{howard2017mobilenets}, and EfficientNet \cite{tan2019efficientnet}. The overall pipeline of the training process for those models is shown in Fig. \ref{fig:workflow}. 
We use the same customisation for all pre-trained models. 
A brief discussion of each pre-trained DL model is presented in the next subsections.

\subsubsection{VGG}
The Visual geometry group (VGG) at Oxford University developed a Convolutional Neural Network (CNN), popularly known as VGG-16, which won the ImageNet \cite{deng2009imagenet} challenge in 2014. It consists of 13 Convolutions, 5 Max pooling and 3 Dense layers. It is named as VGG-16 because it has 16 layers that have the learnable weight parameters \cite{Simonyan15}. 
An extended version of VGG-16 model, which consists 16 Convolution layers, 5 Max-pooling layers and 3 Dense layers, is known as VGG-19.

\subsubsection{ResNet}

The very deep convolutional neural network such as VGG-16 and VGG-19 produced promising results in a large-scale image classification task. However, it is very hard to train a very deep neural network due to a vanishing gradient problem, i.e., the multiplication of small gradient propagated back to the previous layer start vanishing after a certain depth. The researchers aimed to address the vanishing gradient problem 
by introducing the concept of skip connection, which allows skipping some layers in the network. The group of layers in the network that use such skip connection are known as residual blocks (Res-Blocks), which are the core of ResNet architecture \cite{he2016deep}. 
Here, we utilise two ResNet architectures: ResNet-50 and ResNet-101. The ResNet-50 consists of 48 Convolution layers, 1 Max-pooling, and 1 Average pooling layer, whereas ResNet-101 includes 99 Convolution layers, 1 Max-pooling and 1 Average pooling layer. 


\subsubsection{Inception-V3}

The idea of widening the network instead of deepening is implemented in the Inception network, 
by a team of researchers at Google \cite{szegedy2016rethinking}. Inception network architecture uses the four parallel convolutions layers with different kernel sizes at a given depth of network to extract the image feature at different scales before passing them into the next layer. Here, we utilize a 48-layer deep Inception-v3 network.

\subsubsection{InceptionResNet}

With the development of wider and deeper architecture with residual connections such as Inception, \cite{szegedy2016rethinking} and ResNet \cite{he2016deep} network, researchers exploited the benefit of combining the Inception architecture with the residual connections, and established a novel model called InceptionResNet.
We utilise the InceptionResNetV2 network
, which consists of 449 layers including Convolutions layers, Pooling layers, Batch normalization layers and so on.

\subsubsection{Xception}

It is an extreme version of the Inception network developed by Google in 2017 \cite{chollet2017xception}. The main idea implemented in Xception is to make the Convolutions operation more efficient in Inception blocks. This was achieved with modified depth-wise separable convolution, which is performed in two steps: point-wise convolution followed by depth-wise convolution. Here, the point-wise convolution changes the dimension and depth-wise convolution represents the channel-wise spatial convolution. 

\subsubsection{MobileNet}

The idea of depth-wise convolution was further exploited in a deep neural network architecture, known as MobileNet \cite{howard2017mobilenets}. 
In this work, we utilise one version of MobileNet architecture, called MobileNetV2 \cite{sandler2018mobilenetv2}.
The MobileNetV2 is an extended version of MobileNetV1, which consists of 1 regular Convolutions layer, 13 depth-wise separable convolutions blocks and 1 regular Convolutions layer, followed by an Average pooling layer. Whereas, MobileNetV2 added the Expand layer, Residual connections and Projection layers in addition to depth-wise Convolution layers known as a Bottleneck residual block.




\subsubsection{DenseNet}

In DenseNet architecture \cite{huang2017densely}, the idea of skip connection was extended to multiple steps instead of one-step direct connections as in ResNet. And, the block designed to use in between such connections is known as Dense block. The main components of DenseNet are connectivity,  and Dense blocks. Each layer in Densenet has a direct connection to its all forward layer, thereby establishing (L+1)/2 connections for L layer. Each Dense block consists of Convolutions layers with the same feature map size but different kernel sizes. In this work, we utilise the DenseNet-121 network, which consists of 120 Convolutions layers and 4 Average pooling layers.

\subsubsection{EfficientNet}

The expansion of CNN on each dimension into width, depth, and resolution was attempted arbitrarily in various deep neural network architectures such as ResNet, DenseNet, Inception, Xception and so on. However, the systematic approach for scaling up the CNN with a fixed set of scaling coefficients was proposed in EfficientNet architecture \cite{tan2019efficientnet}. The network architecture of EfficientNet consists of three blocks: steam, body and final blocks. The steam and final blocks are common in all variants of EfficientNet while the body differs from one version to another. The stem block consists of input, re-scaling, normalization, padding, convolution, batch normalization and activation layer. The body consists of five modules, where each module has depth-wise convolution, batch normalization and activation layers.
In this study, we use three versions of EfficientNet: EfficientNet-B0, EfficientNet-B1 and EfficientNet-B2. The EfficientNet-Bo has 237 layers in total, whereas EfficientNet-B1 and EfficientNet-B2 have 339 layers, excluding the top layer.

\subsection{Implementation}
We implement our proposed model using Keras \cite{chollet2015keras} implemented in Python \cite{python}. 
During the implementation, we tune the parameters as follows.
We first resize each image into 150*150 as suggested by Sitaula et al. \cite{sitaula2021attention}. 
For augmentation, we apply online data augmentation as follows: rescale=1/255, rotation range=50,width shift range=0.2, height shift range=0.2, shear  range=0.25, zoom range=0.1, and channel shift range=20. We set the optimizer as 'Adam', batch size as 16, and initial learning rate as 0.0001. To prevent over-fitting, we utilise the learning rate decay over each epoch coupled with the Early stopping criteria. 

In our study, we design random five folds (5-cross validation), where each fold contains 70/30 for train/test ratio and report the average performance.

\begin{figure}[tbp]
	\centering
	\includegraphics[height=300mm, width=0.65\textwidth,keepaspectratio]{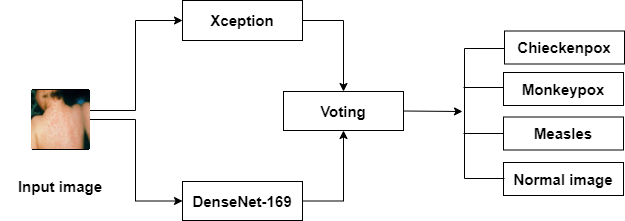}
	\caption{
		Ensemble method between Xception and DenseNet-169 DL models. Note that the Voting block refers to max-voting.
 	} 
	\label{fig:ensemble}
\end{figure}

\begin{figure}[tbp]
	\centering
	\includegraphics[height=200mm, width=0.65\textwidth,keepaspectratio]{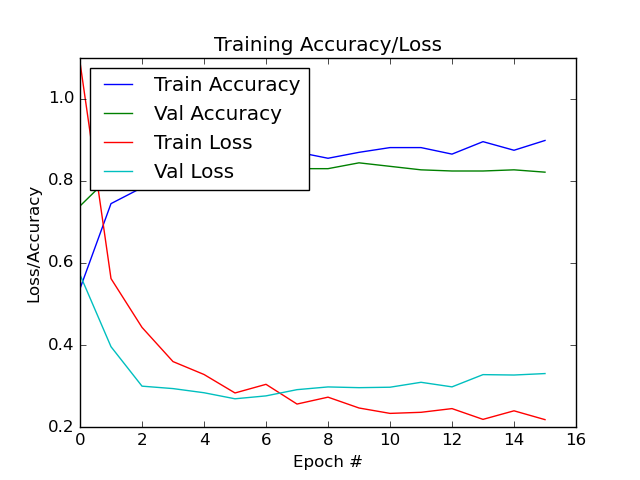}
	\caption{
		Sample train/test plot (fold 1) for accuracy and loss obtained from the fine-tuned Xception DL model.
 	} 
	\label{fig:train_test}
\end{figure}

\subsection{Ensemble approach}

To ensemble the multiple DL models, we extract the probabilistic values from each fine-tuned pre-trained model and perform the majority voting approach (refer to Fig. \ref{fig:ensemble}). Each of our fine-tuned DL models shows the best-fit to learn the optimal features during the training and testing process (see Fig. \ref{fig:train_test}). 

In this study, we choose two best-performing fine-tuned models: Xception and DenseNet-169 based on the empirical study (see Sec. \ref{model_selection}). 
Let us assume that the Xception model produces a probabilistic output vector as $X$ and DenseNet-169 provides a probabilistic output vector as $D$ with a size equal to the number of classes. 

\begin{equation}
    X=Xception(I),
    \label{eq:1}
\end{equation}

\begin{equation}
    D=DenseNet-169(I),
    \label{eq:2}
\end{equation}

\begin{equation}
    C=\operatorname*{arg\,max}_c [X,D],
    \label{eq:3}
\end{equation}
where $I$ is the input image to be classified and C gives us the index of highest majority value corresponding to the particular class $c$.

\begin{table}[b]
    \centering
        \caption{Comparison of pre-trained DL models and ensemble approach using averaged Precision, Recall, F1-score, and Accuracy over 5 different folds. }
    \begin{tabular}{l|p{1cm}|p{1cm}|p{1cm}|p{1cm}}
    \toprule
    Methods&P (\%)&R (\%)&F(\%)&A (\%)\\
    \midrule
     VGG-16&80.18 &79.17 &79.01 & 82.22\\
     VGG-19&81.84&81.90&81.03&82.94\\
     ResNet-50&82.81&82.94 &82.82 &84.87\\
     ResNet-101&82.69&81.88 &82.02 &84.98\\
     IncepResNetv2&83.90&83.44 & 83.62&85.43\\
     MobileNetV2&82.85&81.17 &80.98 &84.87\\
     InceptionV3&82.51&82.30& 82.16 &84.53 \\
     Xception&85.01 & 85.14&  85.02& 86.51\\
     EfficientNet-B0&81.60&81.34&81.40&83.96\\
     EfficientNet-B1&83.69&84.03&83.61&85.09\\
     EfficientNet-B2&82.06&82.67&82.07&83.51\\
     DenseNet-121&83.12&83.00&82.25&84.24\\
     DenseNet-169&84.07&83.74&83.83&86.06\\
     \bottomrule
     Ensemble approach&\bf 85.44&\bf 85.47&\bf 85.40&\bf 87.13\\
     \bottomrule
    \end{tabular}
    \label{tab:comparative_study}
\end{table}

\section{Results and discussion}
\label{results_discussion}

\subsection{Comparative study of DL models}
\label{sota_comparison}

We compare the proposed approach with the off-the-shelf pre-trained DL models based on the standard evaluation measures on this dataset. The results are presented in Table \ref{tab:comparative_study}. Note that the reported results are the averaged performance over 5 different folds (5-fold cross validation).

From Table \ref{tab:comparative_study}, we notice that the Xception is the second-best performing method among all contenders, whereas it is the best method among 13 pre-trained DL methods (Precision: 85.01\%, Recall: 85.14\%, F1-score: 85.02\%, and Accuracy: 86.51\%). We believe that this is because of the Xception's higher ability to extract the discriminating information from the virus images with the help of its point-wise and depth-wise convolution. 
In addition, our proposed ensemble method is the best-performing method among all contenders with an accuracy of 87.13\%. In terms of other performance metrics such as Precision, Recall and F1-score, we observe that it imparts 85.44\% of Precision, 85.47\% of Recall and 85.40\% of F1-score. This shows that it is 0.43\% higher in Precision, 0.33\% higher in Recall, and 0.38\% higher in F1-score than the second-best performing method (Xception). Furthermore, it imparts 5.26\%, 6.3\%, and 6.39\% higher Precision, Recall, and F1-score, respectively than the least-performing DL method (VGG-16). Such improvement in the result is because of the decision fusion, which helps fuse the decision outcomes from different DL models as the final decision.

In summary, our proposed common custom layers are appropriate for fine-tuning all 13 pre-trained DL models to achieve optimal accuracy on this dataset. This is shown not only from the best-fit train/test graph but also from the overall performance (from 80.18\% to 84.07\% for Precision, from 79.17\% to 83.74\% for Recall, from 70.01\% to 83.83\% for F1-score and from 82.22\% to 86.06\% for Accuracy). Similarly, the majority voting approach has also been an interesting option in ensemble learning to exploit the highest decision for the optimal end classification result.

\begin{table}
    \centering
     \caption{Performance comparison of DL different models combination using Precision, Recall, F1-score and Accuracy. Note that M1=Xception, M2=DenseNet-169, M3=IncepResNetv2, M4=EfficientNet-B1, and M5=ResNet-101.
     }
    \begin{tabular}{l|p{1cm}|p{1cm}|p{1cm}|p{1cm}}
    \toprule
    Model ensemble&P (\%)&R (\%)&F (\%)&A (\%)\\
    \midrule
 \{M1, M2, M3, M4, M5\}&84.03&83.26&83.50&85.83\\
 \{M1, M2, M3, M4\}&84.63&84.03&84.24&86.23\\
 \{M1, M2, M3\}&84.55&84.00&84.20&86.17\\
 \bottomrule
 \{M1,M2\}&\bf 85.44&\bf 85.47&\bf 85.40&\bf 87.13\\
     \bottomrule
    \end{tabular}
    \label{tab:model_selection}
\end{table}

\begin{figure*}
\begin{center}
 \subfigure[]{\includegraphics[width=55mm, height=55mm,keepaspectratio]{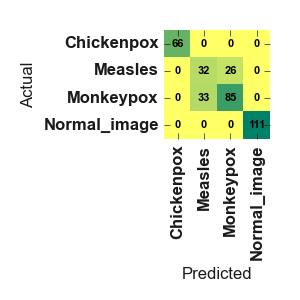}}
 \subfigure[]{\includegraphics[width=55mm, height=55mm,keepaspectratio]{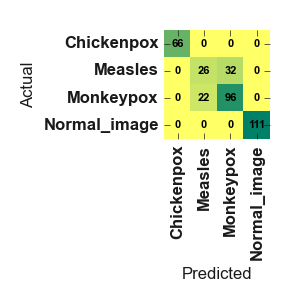}}
 \subfigure[]{\includegraphics[width=55mm, height=55mm,keepaspectratio]{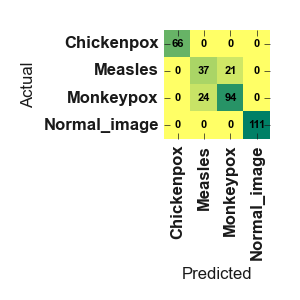}}
 \subfigure[]{\includegraphics[width=55mm, height=55mm,keepaspectratio]{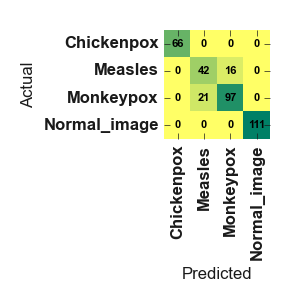}}
 \subfigure[]{\includegraphics[width=55mm, height=55mm,keepaspectratio]{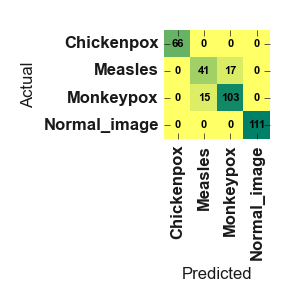}}
  \caption{Confusion matrix of results obtained for five folds from (a) to (e)  from the best performing Xception DL model in our study.}
  \label{fig:confusion_matrix}
 \end{center}
  \end{figure*}

\subsection{Candidate model selection for ensemble}
\label{model_selection}

We select only those DL models that provide optimal performance in our study. For this, we select the top-5 models and their combinations for decision fusion. The detailed results are presented in Table \ref{tab:model_selection}. 
From Table \ref{tab:model_selection}, we observe that the combination of two models (Xception as M1 and DenseNet-169 as M2) provides us with the best performance compared to other combinations. 

\begin{table}[b]
    \centering
     \caption{{Performance comparison of the proposed with the state-of-the-art methods using Precision, Recall, F1-score and Accuracy. }
     }
    \begin{tabular}{l|p{1cm}|p{1cm}|p{1cm}|p{1cm}}
    \toprule
    Methods&P (\%)&R (\%)&F (\%)&A (\%)\\
    \midrule
     CNN-LSTM, 2020 \cite{islam2020combined}&80.40&80.60&79.60&79.60\\
     BoDVW, 2021 \cite{sitaula2021new} &63.20&64.20&61.00&73.80\\
     AVGG, 2021 \cite{sitaula2021attention}&81.60&81.60&80.60&81.80\\
     MBoDVW, 2021 \cite{sitaula2021fusion}&64.60&68.40&65.40&76.80\\
 \bottomrule
 Ensemble approach&\bf 85.44&\bf 85.47&\bf 85.40&\bf 87.13\\
     \bottomrule
    \end{tabular}
    \label{tab:sota_comparison}
\end{table}

\subsection{Comparative study with state-of-the-art methods}
\label{sota_comparision}

{
Although there are no such well-established published state-of-the-art methods for the Monkeypox virus detection, we compare our proposed model with some of the closely related methods that have been used for COVID-19 detection. For this, we utilise four popular well-established DL-based methods: deep bag of words (BoDVW) \cite{sitaula2021new}, multi-scale deep bag of deep visual words (MBoDVW) \cite{sitaula2021fusion}, attention-based VGG (AVGG) \cite{sitaula2021attention} and convolutional neural network with long short term memory (CNN-LSTM) \cite{islam2020combined}. We try our best to select the optimal hyperparameters from the corresponding papers.
The results are presented in Table \ref{tab:sota_comparison}, which show that the proposed method is superior to the state-of-the-art methods in terms of well-established evaluation measures. From this result, we believe our method is appropriate to the Monkeypox virus detection problem, whereas the contender methods are based on chest X-ray images and appropriate to COVID-19 detection problems.}

\subsection{Explainability}
\label{explainability}

\begin{figure}[tbp]
	\centering
	\includegraphics[height=200mm, width=0.65\textwidth,keepaspectratio]{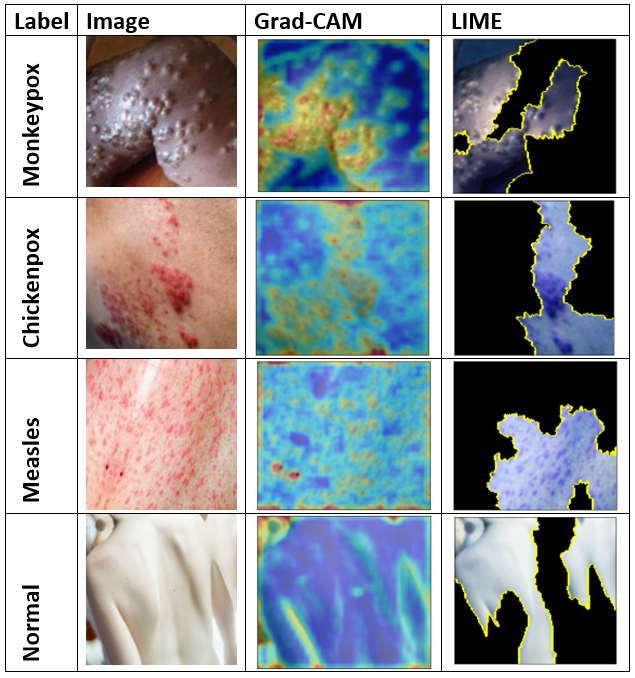}
	\caption{Visualisation based on Grad-CAM and LIME for the Xception model.} 
	\label{fig:explainability}
\end{figure}

We show the explainability of the DL model using  the Gradient-weighted Class Activation Mapping (Grad-CAM) \cite{selvaraju2017grad} and  Local Interpretable Model-Agnostic Explanations (LIME) \cite{lime} visualisation techniques. For this, we use the Xception model, which is the best-performing model, on the Monkeypox dataset. The outputs are presented in Fig. \ref{fig:explainability}. The Grad-CAM measures the gradient of the output feature map of a selected layer of the network, whereas the LIME is a local model-independent approach to generate the interpretation for a specific case, which transforms the input data into a series of interpretable local representations.

From Fig. \ref{fig:explainability}, we notice that the outputs obtained from the Xception model is able to detect the discriminating regions clearly for the classification. For instance, the Grad-CAM is able to show the virus-infected regions in yellow or dark yellow color and the LIME is able to encircle the potentially infected regions with its superpixel on the map. Note that for the Grad-CAM, we set '$block14\_sepconv2\_act$' layer from the Xception DL model. And, for the LIME, we set the number of features as 5, the number of samples as 1000, and top labels as 4 for the Xception DL model.

\subsection{Class-wise study}
\label{ablative}
We study the class-wise performance of the ensemble approach using the confusion matrix, which is shown in Fig. \ref{fig:confusion_matrix}. From Fig. \ref{fig:confusion_matrix}, different confusion matrices for all five-folds show that our ensemble approach is able to discriminate the images clearly into four different classes. More specifically, our method is able to highly discriminate the chickenpox and normal images compared to measles and monkeypox virus.%
{Furthermore, all instances of chickenpox (66)  and normal images (111) in the test set are recognised correctly by the proposed model,  whereas it is still not perfect to discriminate measles and monkeypox viruses in fold 1 (a).}
{
This might be due to the similar features identified by the backbone CNN for two classes: measles and Monkeypox viruses as seen in Grad-CAM visualization \ref{fig:explainability}.
}

\section{Conclusion and future works}
\label{conclusion_futureworks}

In this paper, we compared 13 different pre-trained DL models with the help of transfer learning on the monkeypox dataset. With the help of such comparison using well-established evaluation measures, we identified the best-performing DL models to ensemble them for overall performance improvement. The evaluation result shows that the ensemble approach provides the highest performance (Precision: 85.44\%; Recall: 85.47\%; F1-score: 85.40\%; and Accuracy: 87.13\%) during the detection of the Monkeypox virus. Also, the Xception DL model provides the second-best performance (Precision: 85.01\%; Recall: 85.14\%; F1-score: 85.02\%; and Accuracy: 86.51\%)). 

There are two major limitations of our work. First, the dataset size is comparatively smaller, so addition of more data could improve the performance further. Second, our AI approach is based on pre-trained DL models, which could be a problem if we would like to deploy them in a memory-constrained setting. So, the design of novel lightweight DL models could be an interesting work to let it work on a limited resource. 

\section{Declarations}
\subsection{Ethical Approval and Consent to participate}
Authors declares that no ethical approval is required as the data used in this work are publicly available.
\subsection{Human and Animal Ethics}
Not applicable.
\subsection{Consent for publication}
Not applicable.
\subsection{Availability of supporting data} 
All data are publicly available.
\subsection{Competing interests} 
The authors have no competing interests to declare that are relevant to the content of this article.
\subsection{Funding}
No funding was received for conducting this study.
\subsubsection{Authors' contributions}
C. Sitaula conceived an idea and simulated. C. Sitaula and TB. Shahi revised the manuscript and proofread.
\subsection{Acknowledgements} 
The authors would like to thank those researchers, who created dataset related to Monkeypox detection. 
\subsection{Authors' information} 
C. Sitaula\\
Department of Electrical and Computer Systems Engineering,\\
Monash University\\
Wellignton Rd, VIC 3800 \\
\\TB. Shahi \\
School of Engineering and Technology\\
Central Queensland University \\
Norman Garden,QLD, 4701, Australia\\
and \\
Central Department of Computer Science and IT \\
Tribhuvan University\\
TU Rd, Kirtipur 44618, Kathmandu, Nepal \\              




\bibliographystyle{spbasic}
\bibliography{sample_library}

\begin{thebibliography}{35}
\providecommand{\natexlab}[1]{#1}
\providecommand{\url}[1]{{#1}}
\providecommand{\urlprefix}{URL }
\expandafter\ifx\csname urlstyle\endcsname\relax
  \providecommand{\doi}[1]{DOI~\discretionary{}{}{}#1}\else
  \providecommand{\doi}{DOI~\discretionary{}{}{}\begingroup
  \urlstyle{rm}\Url}\fi
\providecommand{\eprint}[2][]{\url{#2}}

\bibitem[{Ahsan et~al.(2022{\natexlab{a}})Ahsan, Uddin, Farjana, Sakib, Momin,
  and Luna}]{ahsan2022image}
Ahsan MM, Uddin MR, Farjana M, Sakib AN, Momin KA, Luna SA (2022{\natexlab{a}})
  Image data collection and implementation of deep learning-based model in
  detecting monkeypox disease using modified vgg16. arXiv preprint
  arXiv:220601862

\bibitem[{Ahsan et~al.(2022{\natexlab{b}})Ahsan, Uddin, and
  Luna}]{ahsan2022monkeypox}
Ahsan MM, Uddin MR, Luna SA (2022{\natexlab{b}}) Monkeypox image data
  collection. arXiv preprint arXiv:220601774

\bibitem[{Bhatt et~al.(2021)Bhatt, Kumar, Pande, Malik, Khamparia, and
  Gupta}]{Bhatt2021}
Bhatt T, Kumar V, Pande S, Malik R, Khamparia A, Gupta D (2021) A Review on
  COVID-19, Springer International Publishing, chap~2

\bibitem[{Breman et~al.(1980)Breman, Steniowski, Zanotto, Gromyko, Arita
  et~al.}]{breman1980human}
Breman JG, Steniowski M, Zanotto E, Gromyko A, Arita I, et~al. (1980) Human
  monkeypox, 1970-79. Bulletin of the World Health Organization 58(2):165

\bibitem[{Chollet(2017)}]{chollet2017xception}
Chollet F (2017) Xception: Deep learning with depthwise separable convolutions.
  In: Proceedings of the IEEE conference on computer vision and pattern
  recognition, pp 1251--1258

\bibitem[{Chollet et~al.(2015)}]{chollet2015keras}
Chollet F, et~al. (2015) Keras. \url{https://github.com/fchollet/keras}

\bibitem[{Deng et~al.(2009)Deng, Dong, Socher, Li, Li, and
  Fei-Fei}]{deng2009imagenet}
Deng J, Dong W, Socher R, Li LJ, Li K, Fei-Fei L (2009) Imagenet: A large-scale
  hierarchical image database. In: 2009 IEEE conference on computer vision and
  pattern recognition, Ieee, pp 248--255

\bibitem[{Glock et~al.(2021)Glock, Napier, Gary, Gupta, Gigante, Schaffner, and
  Wang}]{glock2021measles}
Glock K, Napier C, Gary T, Gupta V, Gigante J, Schaffner W, Wang Q (2021)
  Measles rash identification using transfer learning and deep convolutional
  neural networks. In: 2021 IEEE International Conference on Big Data (Big
  Data), IEEE, pp 3905--3910

\bibitem[{He et~al.(2016)He, Zhang, Ren, and Sun}]{he2016deep}
He K, Zhang X, Ren S, Sun J (2016) Deep residual learning for image
  recognition. In: Proceedings of the IEEE conference on computer vision and
  pattern recognition, pp 770--778

\bibitem[{Howard et~al.(2017)Howard, Zhu, Chen, Kalenichenko, Wang, Weyand,
  Andreetto, and Adam}]{howard2017mobilenets}
Howard AG, Zhu M, Chen B, Kalenichenko D, Wang W, Weyand T, Andreetto M, Adam H
  (2017) Mobilenets: Efficient convolutional neural networks for mobile vision
  applications. arXiv preprint arXiv:170404861

\bibitem[{Huang et~al.(2017)Huang, Liu, Van Der~Maaten, and
  Weinberger}]{huang2017densely}
Huang G, Liu Z, Van Der~Maaten L, Weinberger KQ (2017) Densely connected
  convolutional networks. In: Proceedings of the IEEE conference on computer
  vision and pattern recognition, pp 4700--4708

\bibitem[{Islam et~al.(2020)Islam, Islam, and Asraf}]{islam2020combined}
Islam MZ, Islam MM, Asraf A (2020) A combined deep cnn-lstm network for the
  detection of novel coronavirus (covid-19) using x-ray images. Informatics in
  medicine unlocked 20:100412

\bibitem[{Lara and Vel{\'a}squez(2022)}]{lara2022low}
Lara JVM, Vel{\'a}squez RMA (2022) Low-cost image analysis with convolutional
  neural network for herpes zoster. Biomedical Signal Processing and Control
  71:103250

\bibitem[{Madhavan et~al.(2021)Madhavan, Khamparia, Gupta, Pande, Tiwari, and
  Hossain}]{madhavan2021res}
Madhavan MV, Khamparia A, Gupta D, Pande S, Tiwari P, Hossain MS (2021)
  Res-covnet: An internet of medical health things driven covid-19 framework
  using transfer learning. Neural Computing and Applications pp 1--14

\bibitem[{Nolen et~al.(2016)Nolen, Osadebe, Katomba, Likofata, Mukadi, Monroe,
  Doty, Hughes, Kabamba, Malekani et~al.}]{nolen2016extended}
Nolen LD, Osadebe L, Katomba J, Likofata J, Mukadi D, Monroe B, Doty J, Hughes
  CM, Kabamba J, Malekani J, et~al. (2016) Extended human-to-human transmission
  during a monkeypox outbreak in the democratic republic of the congo. Emerging
  infectious diseases 22(6):1014

\bibitem[{Pan and Yang(2009)}]{pan2009survey}
Pan SJ, Yang Q (2009) A survey on transfer learning. IEEE Transactions on
  knowledge and data engineering 22(10):1345--1359

\bibitem[{Reynolds et~al.(2013)Reynolds, Emerson, Pukuta, Karhemere, Muyembe,
  Bikindou, McCollum, Moses, Wilkins, Zhao et~al.}]{reynolds2013detection}
Reynolds MG, Emerson GL, Pukuta E, Karhemere S, Muyembe JJ, Bikindou A,
  McCollum AM, Moses C, Wilkins K, Zhao H, et~al. (2013) Detection of human
  monkeypox in the republic of the congo following intensive community
  education. The American Journal of Tropical Medicine and Hygiene 88(5):982

\bibitem[{Ribeiro et~al.(2016)Ribeiro, Singh, and Guestrin}]{lime}
Ribeiro MT, Singh S, Guestrin C (2016) "why should {I} trust you?": Explaining
  the predictions of any classifier. In: Proceedings of the 22nd {ACM} {SIGKDD}
  International Conference on Knowledge Discovery and Data Mining, San
  Francisco, CA, USA, August 13-17, 2016, pp 1135--1144

\bibitem[{Rossum(1995)}]{python}
Rossum G (1995) Python reference manual. Tech. rep., Amsterdam, The Netherlands

\bibitem[{Sandeep et~al.(2022)Sandeep, Vishal, Shamanth, and
  Chethan}]{sandeep2022diagnosis}
Sandeep R, Vishal K, Shamanth M, Chethan K (2022) Diagnosis of visible diseases
  using cnns. In: Proceedings of International Conference on Communication and
  Artificial Intelligence, Springer, pp 459--468

\bibitem[{Sandler et~al.(2018)Sandler, Howard, Zhu, Zhmoginov, and
  Chen}]{sandler2018mobilenetv2}
Sandler M, Howard A, Zhu M, Zhmoginov A, Chen LC (2018) Mobilenetv2: Inverted
  residuals and linear bottlenecks. In: Proceedings of the IEEE conference on
  computer vision and pattern recognition, pp 4510--4520

\bibitem[{Sarumi(2020)}]{sarumi2020machine}
Sarumi OA (2020) Machine learning-based big data analytics framework for ebola
  outbreak surveillance. In: International Conference on Intelligent Systems
  Design and Applications, Springer, pp 580--589

\bibitem[{Selvaraju et~al.(2017)Selvaraju, Cogswell, Das, Vedantam, Parikh, and
  Batra}]{selvaraju2017grad}
Selvaraju RR, Cogswell M, Das A, Vedantam R, Parikh D, Batra D (2017) Grad-cam:
  Visual explanations from deep networks via gradient-based localization. In:
  Proceedings of the IEEE international conference on computer vision, pp
  618--626

\bibitem[{Shahi et~al.(2022{\natexlab{a}})Shahi, Sitaula, and
  Paudel}]{shahi2022hybrid}
Shahi T, Sitaula C, Paudel N (2022{\natexlab{a}}) A hybrid feature extraction
  method for nepali covid-19-related tweets classification. Computational
  Intelligence and Neuroscience 2022

\bibitem[{Shahi et~al.(2022{\natexlab{b}})Shahi, Sitaula, Neupane, and
  Guo}]{shahi2022fruit}
Shahi TB, Sitaula C, Neupane A, Guo W (2022{\natexlab{b}}) Fruit classification
  using attention-based mobilenetv2 for industrial applications. Plos one
  17(2):e0264586

\bibitem[{Simonyan and Zisserman(2015)}]{Simonyan15}
Simonyan K, Zisserman A (2015) Very deep convolutional networks for large-scale
  image recognition. In: International Conference on Learning Representations

\bibitem[{Sitaula and Aryal(2021)}]{sitaula2021new}
Sitaula C, Aryal S (2021) New bag of deep visual words based features to
  classify chest x-ray images for covid-19 diagnosis. Health information
  science and systems 9(1):1--12

\bibitem[{Sitaula and Hossain(2021)}]{sitaula2021attention}
Sitaula C, Hossain MB (2021) Attention-based vgg-16 model for covid-19 chest
  x-ray image classification. Applied Intelligence 51(5):2850--2863

\bibitem[{Sitaula et~al.(2021{\natexlab{a}})Sitaula, Basnet, Mainali, and
  Shahi}]{sitaula2021deep}
Sitaula C, Basnet A, Mainali A, Shahi TB (2021{\natexlab{a}}) Deep
  learning-based methods for sentiment analysis on nepali covid-19-related
  tweets. Computational Intelligence and Neuroscience 2021

\bibitem[{Sitaula et~al.(2021{\natexlab{b}})Sitaula, Shahi, Aryal, and
  Marzbanrad}]{sitaula2021fusion}
Sitaula C, Shahi TB, Aryal S, Marzbanrad F (2021{\natexlab{b}}) Fusion of
  multi-scale bag of deep visual words features of chest x-ray images to detect
  covid-19 infection. Scientific reports 11(1):1--12

\bibitem[{Szegedy et~al.(2016)Szegedy, Vanhoucke, Ioffe, Shlens, and
  Wojna}]{szegedy2016rethinking}
Szegedy C, Vanhoucke V, Ioffe S, Shlens J, Wojna Z (2016) Rethinking the
  inception architecture for computer vision. In: Proceedings of the IEEE
  conference on computer vision and pattern recognition, pp 2818--2826

\bibitem[{Tan and Le(2019)}]{tan2019efficientnet}
Tan M, Le Q (2019) Efficientnet: Rethinking model scaling for convolutional
  neural networks. In: International conference on machine learning, PMLR, pp
  6105--6114

\bibitem[{Unnikrishnan et~al.(2022)Unnikrishnan, Gontu, Khwairakpam, and
  Sagar}]{Unnikrishnan2022}
Unnikrishnan M, Gontu HL, Khwairakpam BS, Sagar P (2022) Detection of covid
  from chest x-rays using gan. EPRA International Journal of Research and
  Development (IJRD) 7:166–175,
  \urlprefix\url{http://www.eprajournals.net/index.php/IJRD/article/view/453}

\bibitem[{{World health organization}(2022)}]{who}
{World health organization} (2022) Multi-country monkeypox outbreak: situation
  update.
  \url{https://www.who.int/emergencies/disease-outbreak-news/item/2022-DON396},
  (Accessed: 2022-06-30)

\bibitem[{Yadav et~al.(2022)Yadav, Alfayeed, Khamparia, Pandey, Thanh, and
  Pande}]{yadav2022hsv}
Yadav N, Alfayeed SM, Khamparia A, Pandey B, Thanh DN, Pande S (2022) Hsv
  model-based segmentation driven facial acne detection using deep learning.
  Expert Systems 39(3):e12760

\end{thebibliography}
\end{document}